\documentclass[draft,11pt]{article}
\usepackage{amsmath,amssymb,epsfig,graphicx,,color,epsf}
\newtheorem{theorem}{Theorem}
\newtheorem{proposition}{Proposition}
\newtheorem{lemma}{Lemma}

\newtheorem{definition}{Definition}

\newcommand{\begsection}[1]{\setcounter{equation}{0}\section{#1}}

\def\H{{\mathcal H}}

\def\vf{\varphi}

\def\psibx{\overline{\psi(x)}}
\def\psiby{\overline{\psi(y)}}
\def\As{{\cal A}_{\sigma,p}}
\def\R{\mathbb R}

\def\N{\mathbb N}

\def\A{{\mathcal A}}

\def\Sc{Schr\"o\-din\-ger}
\def\hp{{\hbar}}
\def\la{\langle}
\def\be{\begin{equation}}
\def\ee{\end{equation}}
\def\ra{\rangle}
\def\ds{\displaystyle}

\def\ep{\epsilon}

\def\limN{\lim_{N\to\infty}}

\def\be{\begin{equation}}
\def\ee{\end{equation}}
\def\ap{a^{(p)}}
\def\alp{\alpha^{(p)}}

\def\w2{w^{(2)}}

\def\apn{A^{(p)}_N}
\begin{document}
\baselineskip=20pt
\date{15 February 2005}
\begin{center}
{\large\bf 
MEAN-FIELD- AND CLASSICAL LIMIT OF MANY-BODY SCHR\"ODINGER DYNAMICS FOR BOSONS
}
\end{center}
\vskip 13pt
\begin{center}
J\"urg Fr\"ohlich\footnote{Theoretische Physik, ETH Z\"urich, Switzerland. (juerg@itp.phys.ethz.ch)},
Sandro Graffi\footnote{Dipartimento di Matematica,  Universit\`{a} di Bologna,  Italy.
(graffi@dm.unibo.it)} ,
 Simon Schwarz\footnote{Theoretische Physik, ETH Z\"urich, Switzerland.
(sschwarz@itp.phys.ethz.ch)}\end{center}
\begin{abstract}
\noindent
We present a new proof of the convergence  of the $N-$particle
\Sc\ dynamics for bosons towards the dynamics  generated by the Hartree equation in the mean-field limit. For a
restricted class of two-body interactions, we obtain convergence estimates uniform 
in
$\hbar$, up to an exponentially small remainder. For $\hbar = 0$, the
classical dynamics in the mean-field limit is given by the Vlasov equation.
\end{abstract}
\vskip 1cm   
\date{\today} 
\begsection{Introduction and statement of results}
\setcounter{equation}{0}%
\setcounter{theorem}{0}%
\setcounter{proposition}{0}%
\setcounter{corollary}{0}%
\setcounter{definition}{0}%

Consider the \Sc\ operator
\begin{eqnarray} \label{hn}
H_N&=&H_{N}^0+W_{N}
\\
H_{N}^{0}&=&-\sum_{i=1}^N\frac{\hbar^2}{2}\Delta_i,
\quad W_{N}=\frac1{N}\sum_{i<j}^N\,w(x_i-x_j)
\label{h0}
\end{eqnarray}
where $w$ is a two-body potential {\it independent} of $N$. The operator 
$H_N$ acts on $\H^{(N)}:=L^2_S(\R^{3N})$, the totally symmetric part
of $L^2(\R^{3N})$, which is the Hilbert space of pure state vectors for a system of $N$ nonrelativistic bosons.
We propose to study the dynamics  
described by the $N-$body \Sc\ equation
\be
\label{sc}
i\hbar\partial_t\,\Psi_N(t)=H_N\Psi_N(t),
\ee
for an inital condition 
$\Psi_{N}(t=0)=\Psi_{N,0}\in L_{S}^2(\R^{3N})$. 
Under assumptions specified below, 
$H_N$, defined on the symmetrized 
Sobolev space $H^2_S(\R^{3N})$, 
is a self-adjoint operator.
Hence the unitary group $\ds U_N(t)=e^{-iH_N t/\hbar}$, $t\in\R$, exists.  
Let $p\leq N$, and
let $a^{(p)}$ be a bounded operator on $L^2_S(\R^{3p})$.  
It  defines  an operator $A^{(p)}_N$ acting  on $\H^{(N)}$ 
 in the following way:
\begin{align}
\nonumber
&
(A^{(p)}_N\Psi)(x_1,\ldots,x_N)=\frac{N(N-1)\cdots (N-p+1)}{N^p}
(P_{S}a^{(p)} \otimes I^{(N-p)} P_{S} \Psi)(x_1,\ldots,x_N), 
\\
\label{A_N}
&
\Psi(x_1,\ldots,x_N)\in L^2_S(\R^{3N}), 
\end{align}
where $P_{S}$ is the projection onto the symmetric subspace  $L^{2}_S(\R^{3N})$ of $L^{2}(\R^{3N})$. 
The operator $A^{(p)}_{N}$ may be viewed 
as an  operator acting on $p$ particles; 
the numerator on the right side of (\ref{A_N}) is a combinatorial factor motivated by "second quantization";  the denominator is the correct scaling factor 
to take the $N\to\infty $ limit. 

We are interested in the asymptotics of certain expectation values of the 
Heisenberg-picture operators $\ds   e^{iH_Nt/\hbar}A^{(p)}_Ne^{-iH_Nt/\hbar}$, as $N\to\infty$. If $H_{N}$ is chosen as in (\ref{hn}), (\ref{h0}), and $A^{(p)}_{N}$ is chosen as in (\ref{A_N}), the limit $N\rightarrow\infty$ is the usual {\it mean-field limit}; see \cite{He,Sp}.

Our first main result is the following
\begin{theorem}. 
\label{mainth}
Let $\hbar >0$ and $t\geq 0$ be fixed, and let $w\in L^\infty(\R^3)$. If  
$\Psi_{N,0}(x_1,\ldots,x_N)=\psi(x_1)\cdots\psi(x_N)$ is a normalized "coherent" (i.e., product) 
initial state, then
\begin{eqnarray}
\label{t11}
 \lim_{N\to\infty}\langle\Psi_{N,0},e^{iH_Nt/\hbar}A^{(p)}_Ne^{-iH_Nt/\hbar}\Psi_{N,0}\rangle  = \qquad 
\\
\nonumber
\limN\langle\Psi_{N,t},A^{(p)}_N\Psi_{N,t}\rangle 
 =\la\Psi_{p,t},a^{(p)}\Psi_{p,t}\ra =:a^{(p)}(\psi_{t})
\end{eqnarray}
Here $\Psi_{N,t}$ is again a coherent state, i.e., $\Psi_{N,t}(x_1,\ldots,x_N)=
\psi_t(x_1)\cdots\psi_t(x_N)$, and $\Psi_{p,t} = \Psi_{N=p,t}$, where $\psi_t$ is a solution of the Hartree equation
\be
\label{Hartree}
i\hbar\partial_t\psi_{t}=-\frac{\hbar^2}{2}\Delta\psi_{t}+(w\ast|\psi_{t}|^{2})\psi_{t}
\ee
with initial condition $\psi_{t=0} = \psi$.
\end{theorem}
{\bf Remarks}
\begin{enumerate}
\item  For large $N$, the quantum evolution $e^{-iH_{N}t/\hbar}\Psi_{N,0}\quad$ can be
 replaced by the {\it nonlinear} single-particle evolution $\Psi_{N,t} (x_{1},\ldots,x_{N})$.
Particle interaction effects are translated into the nonlinearity of this evolution. 
This justifies interpreting the limit $N\rightarrow\infty$ as a mean-field limit.
\item
The corrections to the limit in (\ref{t11}) are $O(1/N)$.
\item Since $\ds \limN   {N(N-1)\cdots (N-p+1)}/{N^p}=1$, the second equality in (\ref{t11}) follows easily from (\ref{A_N}), because
$$
\limN \la\Psi_{N,t},A^{(p)}_N\Psi_{N,t}\ra=
\la\Psi_{p,t}, a^{(p)}\Psi_{p,t}\ra_{L^2(\R^{3p})}\|\psi\|_{L^2(\R^{3})}^{2(N-p)}
$$
\item Theorem \ref{mainth} was first proven in \cite{He}, see also [GiVe]. A new proof was given in \cite{Sp} and
extended to more general classes of two-body potentials, including the Coulomb potential, in \cite{EY}, \cite{BGM},
\cite{BEGMY}. The proof in our paper is quite different. It is inspired by a second-quantization
formalism to be published elsewhere. It enables us to tackle the problem of obtaining
convergence estimates uniform in Planck's constant $\hbar$, as we now proceed to discuss. 
\end{enumerate}
It is well known that, for $W_{N}$ as in (\ref{h0}), 
the classical dynamics of $N$ particles tends to the dynamics defined by the 
Vlasov equation, in the limit $N\to\infty$. More precisely, if $\rho_N$ denotes the empirical distribution,
namely
$$
\rho_N(dx,d\xi;t)=\frac1{N}\sum_{i=1}^N\,\delta(x-x_i(t))\delta(\xi-\xi_i(t))\,dxd\xi
$$
where $(x_1(t),\ldots,x_N(t);\xi_1(t),\ldots,\xi_N(t))$ is a solution of the classical equations of motion, then, in the limit $N\to\infty$, $\rho_{N}$
 tends weakly to $f_t(x,\xi) dx d\xi$, where  $f_t(x,\xi)$ is a  solution of the Vlasov equation:
\begin{eqnarray}
\label{Vlasov}
\partial_{t} f_t & = & -\xi\cdot \nabla_{x} f_t + \nabla_{x} V_{eff} \cdot \nabla_{\xi} f_t\;\\
V_{eff} (x,t) & = &\int w(x-y) f_{t}(y,\xi) dyd\xi\;,
\end{eqnarray}
see \cite{BH}. 
It is natural to ask whether this convergence result is related to that of Theorem 1.1. Our next result provides, under very restrictive assumptions on the two-body interactions, a partial answer to this question. First, we define a restricted class 
of interactions.
For $\sigma > 0$, we define the spaces
\begin{eqnarray}
L^1_{\sigma,p}&:=&\{f\in L^1(\R^{6p})\,|\,e^{\sigma|z|}f\in L^1(\R^{6p})\},
\\
{\cal A}_{\sigma,p}&:=&\{f\in L^1(\R^{6p})\,|\,e^{\sigma|s|}\widehat f \in L^1(\R^{6p})\},
\end{eqnarray}
Here $x_j\in\R^3$, $\xi_j\in\R^3$, $j=1,\ldots,p$, and
\newline
$z:=(X_p,\Xi_p)\in\R^{3p}\times\R^{3p}$; $X_p:=(x_1,\ldots,x_p)$, $\Xi_p:=(\xi_1,\ldots,\xi_p)$, 
$$ 
|z|:=\sum_{j=1}^p(|x_j|+|\xi_j|);
$$ 
$\widehat f(s), s:=(S,\Sigma)\in\R^{3p}\times\R^{3p}$ is the Fourier transform of $f$.

We further denote by
$\Phi^N_t: (X_N;\Xi_N)\mapsto (X_N(t);\Xi_N(t))$ the flow generated by 
$H_N^c$, where $H^c_N$ is the  classical Hamilton function corresponding to the operator $H_N$.
\begin{definition}
 We define by:
 \begin{enumerate}
 \item
\begin{align}
\label{W1}
& W_N^{\Psi_N}(X_N,\Xi_N;t)=
\\
\nonumber
& (2\pi)^{-3N}\int_{\R^{3N}}e^{i\la Y_N,\Xi_N\ra}\Psi_N(X_N+\hbar Y_N/2,t)
\overline{\Psi}_N(X_N-\hbar Y_N/2,t)\,dY_N
\end{align}
the {\rm Wigner distribution} of the $N$-particle normalized wave function $\Psi_N(X_N,t)$;
\item
\begin{eqnarray}
\label{W2}
W_j^{\Psi_N} (X_j,\Xi_j;t)=\int_{\R^{3(N-j)}}W_N^{\Psi_N}
(X_N,\Xi_N;t)\,dX_{N-j}d\Xi_{N-j},
\end{eqnarray}
the $j-${\rm particle Wigner function} ($(N-j)$-marginal distribution of the $N$-particle Wig\-ner
distribution).
\item
\be
\label{W3} 
W(\psi)(x,\xi;t)=(2\pi)^{-3}\int_{\R^{3}}e^{i\la y,\xi\ra}\psi_{t}(x+\hbar
y/2)\overline{\psi_{t}}(x-\hbar  y/2)\,dy, 
\ee
the Wigner distribution of the solution $\psi_{t}(x)$ of the Hartree equation.
\end{enumerate}
\end{definition}
\noindent
Our second main result is
\begin{theorem}.
\label{unif}
Let $w\in {\cal A}_{\sigma,1}$, for some
$\sigma>0$. Let $\Psi_{N,0}$ be a product state. Set $\ep:=\|w\|_\infty \,t$. 
Then, for fixed $p$, there is a constant $C_p>0$ \underline{independent} of $\hbar$ such that, as an equality between
tempered distributions, 
\be
\label{uniff1}
W_p^{\Psi_N}
(X_p,\Xi_p;t)=\prod_{j=1}^pW(\psi)(x_j,\xi_j;t)+\frac{C_p}{N}+O\left(e^{-1/\sqrt{\ep}}\right),
\ee
 as $N\to\infty$.
\end{theorem}
{\bf Remarks}
\begin{enumerate}
\item 
It is known that $W(\psi)(x,\xi;t)$ converges in ${\cal S}^\prime(\R^6)$ to a solution $f_{t}(x,\xi)$
of the Vlasov equation, as $\hbar\to 0$ \cite{NS}. It is also known that 
$$
\limN  W_p^{\Psi_N} (X_p,\Xi_p;t)=\prod_{j=1}^p f_{t}(x_j,\xi_j)
$$
whenever $N\to\infty$ entails $\hbar\to 0$, as in the case of the Kac potentials \cite{NS},\cite{GMP}.
\item
 Result (\ref{uniff1}) shows that, up to an exponentially small error independent  of $\hbar$,
the mean-field convergence towards a single-particle nonlinear dynamics 
holds {\it uniformly} in $\hbar$.
\item The classical limit is equivalent to the limit of heavy particles. We set $\hbar=1$ in
(\ref{h0}), but let the particle mass $m$ become large. We impose the condition that the kinetic energy per particle be independent of $m$, namely $mv^2_i=O(1)$, i.e., $\ds |v_i|=O(1/\sqrt{m})$, for all $i$. This
suggests to rescale time as $t=\sqrt{m}\tau$. Then the Schr\"odinger equation becomes
$$
\frac{i}{\sqrt{m}}\partial_\tau\Psi_N=\sum_{j=1}^N -\frac{\Delta_j}{2m}\Psi_N+
\frac{1}{N}\sum_{i,j=1}^Nw(x_i-x_j)\Psi_N,
$$
which is equivalent to (\ref{hn})-(\ref{sc}), for $\hbar=1/\sqrt{m}$. 
\end{enumerate}
\vskip 1cm\noindent
\begsection{The $N\to\infty$ limit: convergence estimates}
\subsection{Kinematical algebra of "observables"} 
The above systems can be described by a kinematical algebra of operators, the quantum mechanical analogue of the algebra of functions on phase space of a classical system.

Let  $\H^{(p)}:=L^2_S(\R^{3p})$, $0<p<N$, $N\in\N$. Let $\ap$ be a bounded ope\-rator on $\H^{(p)}$, and $\alp(x_1,\ldots,x_p;y_1,\ldots,y_p):=\alp(X_p;Y_p)$ be the tempered distribution kernel in ${\mathcal S}^\prime(\R^{3p}\times\R^{3p})$ associated to $\ap$  by the nuclear theorem:
\be
\label{10}
(\ap\vf^{(p)})(X_p)=\int_{\R^p}\alp (X_p;Y_p)\vf^{(p)}(Y_p)\,dY_p
\ee
where $\ds \vf^{(p)}(Y_p)\in L^2_S(\R^{3p})$. 
Then $(\ap)^\ast$ has the distribution kernel $\overline{\alp(Y_p;X_p)}$.
\par\noindent
To $\ap$ we associate the operator ${A}_N^{(p)}$ on $L^2_S(\R^{3N})$  specified in (\ref{A_N}). Explicitly:
\begin{equation}
\label{11}
{A}_N^{(p)}(\ap)\vf^{(N)}:=\left({N}\atop{p}\right)\frac{p!}{N^p}  P_S
\int_{\R^{3N}}K((\cdot),Y_N)\vf^{(N)}(Y_N)\,dY_N,
\ee
with
\be
K(X_N,Y_N)=\alp(X_p;Y_p)\,\delta(X_{N-p}-Y_{N-p}),\;
\vf^{(N)}(Y_N)\in L^2_S(\R^{3N}) 
\end{equation}
If $\ap$ is bounded on $\H^{(p)}$ then ${A}_N^{(p)}(\ap)$ is bounded on $\H^{(N)}$.  Since $\|P_S\|=1$ and $\ds \left({N}\atop{p}\right)\frac{p!}{N^p}\leq 1$,  we have that
\begin{align}
\nonumber
& \|{A}_N^{(p)}\vf^{(N)}\|_{\H^{(N)}}^2  \leq  \|\ap\|_{\H^{(p)}}^2 \int_{R^{3(N-p)}}\left(\int_{R^{3p}}
|\vf^{(N)}(Y_p;X_{N-p})|^2\,dY_p\right)\,dX_{N-p}
\\
\nonumber
& = \|\ap\|_{\H^{(p)}}^2 \|\vf^{(N)}\|_{\H^{(N)}}^2
\end{align}
We set
\be
\label{14}
\widehat{\mathcal A}_N:=\la {A}_N^{(p)}(\ap)\,|\,\ap\in B(\H^{(p)}),\,p=0,1,2,\ldots\ra \subset B(\H^{(N)}).
\ee
The following statement is easily verified.
\begin{proposition}. 
The map   $\ap\mapsto {A}_N^{(p)}(\ap)$ is linear, and  $({A}_N^{(p)}(\ap))^\ast=
A_N^{(p)}((\ap)^\ast)$, $\ds \|{A}_N^{(p)}(\ap)\|_{B(\H^{(N)})}\leq \|\ap\| _{B(\H^{(p)})}$.  
\end{proposition}
\subsection{The Schwinger-Dyson expansion} 
Given a bounded operator $A^{(p)}_N$ acting on $\H^{(N)}$, $p\leq N$, we denote by $A^{(p)}_{t,N}$ the corresponding Heisenberg-picture operator with respect to the free time evolution $\ds e^{iH_N^0t/\hbar}$, i.e.,
\be
\label{A_t}
A^{(p)}_{t,N} = e^{iH_N^0t/\hbar}A^{(p)}_Ne^{-iH_N^0t/\hbar}
\ee
We further denote by $A^{(p)}_{N}(t)$ the corresponding operator with $H_N^0$ replaced by $H_N$, namely
\be
\label{At}
A^{(p)}_{N}(t):= e^{iH_Nt/\hbar}A^{(p)}_Ne^{-iH_Nt/\hbar}, 
\ee
and by  $A^{(p)}_{I,N}(t,s)$ the two-parameter operator family
\be
\label{I_t}
A^{(p)}_{I,N}(t,s):= e^{iH_Nt/\hbar}e^{-iH_N^0t/\hbar}A^{(p)}_{s,N}e^{iH_N^0t/\hbar}e^{-iH_Nt/\hbar}
\ee
Then we  obviously have 
\be
\label{Ist}
A^{(p)}_{N}(t)=\left.A^{(p)}_{I,N}(t,s)\right|_{s=t}
\ee
We denote $W_{N}$ by $W$.  Iterating the identity:
\be
\label{Duhamel1}
A^{(p)}_{I,N}(t,s)=A^{(p)}_{s,N}+\frac{i}{\hbar}\int_0^t\,e^{iH_N t_1/\hbar}e^{-iH_N^0 t_1/\hbar} [W_{t_1},A^{(p)}_{s,N}]e^{iH_N^0 t/\hbar}e^{-iH_N/\hbar} \,dt_1
\ee
we get that 
\be
A^{(p)}_{I,N}(t,s)=A^{(p)}_{s,N}+\sum_{n=1}^\infty \left(\frac{i}{\hbar}\right)^n\int_0^t\,dt_1\ldots \int_0^{t_{n-1}}\,dt_n\,[W_{t_n},\ldots,[W_{t_1},A_{s,N}^{(p)}]\ldots]
\ee
and finally, setting $s=t$, we obtain the Schwinger-Dyson expansion
\begin{align}
\label{Duhameln}
A^{(p)}_N(t) = A^{(p)}_{t,N}+
\sum_{n=1}^\infty\left(\frac{i}{\hbar})\right)^n
\int_0^t\int_0^{t_1}\cdots\int_0^{t_{{n-1}}}[W_{t_n},\ldots,[W_{t_1},A_{t,N}^{(p)}]\ldots]\,dt_{n}
\ldots dt_1
\end{align}
From now on, we drop the index $N$ in the Heisenberg-picture operators with respect to the free evolution, i.e. we use the abbreviation: $A^{(p)}_{t,N}:=A^{(p)}_{t}$. 
\newline
The boundedness of $A^{(p)}_N$ and of the interactions $W_{t_i}$ implies the boundedness of all multiple 
commutators, with 
\begin{align}
\nonumber 
\frac{1}{\hbar^n}\|[W_{t_n},\ldots,[W_{t_1},A_t^{(p)}]\ldots]\|_{\H^{(N)}} & \leq (2\|W\|_{\H^{(N)}}/\hbar)^n\|A^{(p)}_N\|_{\H^{(N)}}
\\
\nonumber
& \leq (2\|W\|_{\H^{(N)}}/\hbar)^n\|\ap\|_{\H^{(p)}}, 
\end{align}
for $A^{(p)}_N=A^{(p)}_N(\ap)$. By (\ref{h0}), $\ds \|W\|_{\H^{(N)}}\propto N$. Hence, for fixed $N$ and $\hbar$, the series is norm-convergent, 
for all $t\geq 0$. The time integrations yield a factor $\ds \frac{t^n}{n!}$, so that the norm of the series in (\ref{Duhameln}) 
is bounded by $\ds \exp{[2\|W\|_{\H^{(N)}}|t|/\hbar]}\cdot\|a^{(p)}\|_{\H^{(p)}}$. 

These estimates are obviously not adequate to investigate the $N\to\infty$ or the
$\hbar\to 0$ limit, let alone to prove uniformity in $\hbar$.
\subsection{The $N\to\infty$ limit }

We  exploit the structure of the commutators on the right-hand side of (\ref{Duhameln}), the symmetry of wave functions in $L^{2}_{S}(\R^{3N})$, and
the fact that each term in $ A^{(p)}_N$ only acts on $p$ arguments of a wave function, so that many commutators
will vanish. Note that
\begin{equation}
\label{W_t}
W_t=\frac1{N}\sum_{i<j}^Ne^{-iH_N^{0} t/\hbar}w^{ij}e^{iH_N^{0}t/\hbar}=
\frac1{N}\sum_{i<j}^N w_t^{ij},
\end{equation}
where 
\begin{equation}
w_t^{ij}=e^{i(\Delta_i+\Delta_j)t\hbar/2}w^{ij}e^{-i(\Delta_i+\Delta_j)t\hbar/2},
\; w^{ij}=w(x_i-x_j).
\end{equation}
Therefore
\begin{eqnarray}
\label{comm11}
\nonumber
[W_s,A^{(p)}_{N}]&=&\frac1{N}\sum_{i<j}^N[w_s^{ij},A^{(p)}_{N}]=
\\
\nonumber
&=&
\frac1{N}\sum_{i=1}^p\sum_{j=p+1}^NA_N^{(p+1)}([w_s^{ij},\ap])+
\frac1{N}\sum_{i<j}^pA_N^{(p)}([w_s^{ij},\ap])=
\\
&=&
\frac{N-p}{N}\sum_{i=1}^pA_N^{(p+1)}([w_s^{ip+1},\ap])+
\frac1{N}\sum_{i<j}^pA_N^{(p)}([w_s^{ij},\ap])
\end{eqnarray}
 In more
precise terms, the expression
\begin{eqnarray}
\label{comm1}
[W_s,A^{(p)}_{N}]&=&\frac{N-p}{N}\sum_{i=1}^pA^{(p+1)}_{N}([w_s^{ip+1},a^{(p)}])+
\frac1{N}\sum_{i<j}^pA^{(p)}_{N}([w_s^{ij},a^{(p)}]),
\end{eqnarray}
holds as an operator identity on $\H^{(N)}$. In second-quantization language, the first sum on (\ref{comm1}) corresponds to {\it tree graphs}, the second one to {\it loop graphs}. Next, we insert (\ref{comm1}) in (\ref{Duhamel1}) and perform a second step,
but only for the first sum in (\ref{comm1}), leaving the second one unchanged. To
keep our notation compact, it is useful to introduce the notion of {\it tree amplitudes} of
$n-$th order, recursively defined in the following way
\be
\label{treen}
g^{(0,p)}=\ap;\quad
g^{(n;p)}_{t_1,\ldots,t_n}=\frac{i}{\hbar}\sum_{i=1}^{p+n-1}[w_{t_n}^{ip+n},
g^{(n-1;p)}_{t_1,\ldots,t_{n-1}}],\;\; n\geq 1
\ee
Then expression (\ref{comm1}) becomes
$$
\frac{i}{\hbar}[W_s,A^{(p)}_{N}]=\frac{N-p}{N}
A^{(p+1)}_{N}(g^{(1;p)}_{s})+
\frac{i}{N\hbar}\sum_{i<j}^p\apn([w_s^{ij},g^{(0;p)}]).
$$
The first term is $O(1)$, while the second one is of order $p(p-1)/N$ (for fixed $\hbar$) and is
therefore suppressed by a factor $1/N$. Performing $(k-1)$ iterations only for the
{\it tree amplitudes}, we conclude that 
\be
\label{Duhamel3}
e^{iH_Nt/\hbar}A^{(p)}_Ne^{-iH_Nt/\hbar}=
A^{(p)}_{t,N}+B_{t,N}^{(p),k}+\frac1{N}\sum_{n=1}^k Q_{t,N}^{(p),n} +R_{t,N}^{(p),k}, 
\ee
where
\begin{align}
\label{Duhamel4}
&
B_{t,N}^{(p),k} =\sum_{n=1}^{k-1}\,\int_0^t\ldots\int_0^{t_{n-1}}A^{(p+n)}_{N}(g^{(n;p)}_{t_1,\ldots,t_{n}})\,dt_{n}\ldots dt_1 ,
\\
\label{Duhamel5}
& Q_{t,N}^{(p),n}  =
\sum_{j>i=1}^{p+n-1}
\int_0^t\ldots\int_0^{t_{n-1}}
e^{iH_N^0 t_{n}/\hbar}e^{-iH_N t_{n}/\hbar}
H(n-1;p;N)
e^{iH_N t_{n}/\hbar}e^{-iH_N^0 t_{n}/\hbar}
\,dt_{n}\ldots
dt_1,
\\
\label{Duhamel6}
& R^{(p),k}_{t,N}  = 
\int_0^t\ldots\int_0^{t_{k-1}}
e^{iH_N^0 t_{n}/\hbar}
e^{-iH_Nt_{n}/\hbar}H(k,p;N)
e^{iH_N t_{n}/\hbar}e^{-iH_N^0 t_{n}/\hbar}
\,dt_{k}\ldots
dt_1,
\end{align}
with
\be
\label{HPN}
 H(s,p;N):=\sum_{i<j}^{p+s-1}A^{(p+s-1)}_{N}([w^{ij}_{t_s},
g^{(s;p)}_{t_1,\ldots,t_{s-1}}])
\end{equation}
Equation (\ref{Duhamel3}) is most easily verified as follows (think of  $\ds A^{(p+n-1)}_{N}(g^{n-1,p}_{t_1,\ldots,t_{n-1}})$ as a $p+n-1$-particle operator replacing $A_{t,N}^{(p)}$ in (\ref{comm1}) ):
\begin{align*}
& i \frac{[W_{t_n},
A^{(p+n-1)}_{N}(g^{n-1,p}_{t_1,\ldots,t_{n-1}})]}{\hbar}=
\frac{(N-(p+n-1))}{N}
i\sum_{i=1}^{p+n-1}{A^{(p+n)}_{N}([w^{ip+n}_{t_n},
g^{n-1,p}_{t_1,\ldots,t_{n-1}}]}) +
\\
&
\frac{1}{N\hbar}\sum_{j>i=1}^{p+n-1}iA^{(p+n)}_{N}([w^{ij}_{t_n},g^{n-1,p}_{t_1,\ldots,t_{n-1}}])=
A^{(p+n)}_{N}(g^{n,p}_{t_1,\ldots,t_{n}})+
\\
&
\frac{1}{N\hbar}\sum_{j>i=1}^{p+n-1}iA^{(p+n)}_{N}
({[w_{t_n}^{ij},g^{(n-1;p)}_{t_1,\ldots,t_{n-1}}]})
\end{align*}
\subsection{Control of the expansion, small time, $\hbar$ fixed}
First, we prove a bound on the norm of $A^{(p+n)}_{N}(g^{(n;p)}_{t_1,\ldots,t_{n}})$
\be
\label{NormGn}
\|A^{(p+n)}_{N}(g^{(n;p)}_{t_1,\ldots,t_{n}})\|_{\H^{(N)}}\leq \frac{2}{\hbar}(p+n-1)\|w\|_\infty\,
\|A^{(p+n-1)}_{N}(g^{(n-1;p)}_{t_1,\ldots,t_{n-1}})\|_{\H^{(N)}}
\ee
This follows from the unitarity of the free time evolution and the boundedness of the
interactions, $\|w^{ij}\|=\|w\|_\infty$. The bound (\ref{NormGn}) then yields recursively
\begin{eqnarray}
\label{normGn1}
\nonumber
\|A^{(p+n)}_{N}(g^{(n;p)}_{t_1,\ldots,t_{n}})\|_{\H^{(N)}}&\leq & (p+n-1)(p+n-2)\cdots (p+1)p
\left(\frac{2}{\hbar}\|w\|_\infty\right)^n\|A^{(p)}_N\|_{\H^{(N)}}
\\
&\leq &
\frac{(p+n)!}{p!}\left(\frac{2}{\hbar}\|w\|_\infty\right)^n\|\ap\|_{\H^{(p)}},
\end{eqnarray}
independently of all time indices. 

Considering the expansion (\ref{Duhamel3}), we have that
\begin{align}
\nonumber
&
\left\|\sum_{n=1}^\infty\int_0^t\ldots\int_0^{t_{n-1}}
A^{(p+n)}_{N}(g^{(n;p)}_{t_1,\ldots,t_{n}})\,dt_n\cdots dt_1\right\|_{\H^{(N)}} \leq
\\
\label{est1}
&
\sum_{n=1}^\infty\frac{t^n}{n!}\frac{(p+n)!}{p!}(\frac{2}{\hbar}\|w\|_\infty t)^n\|A^{(p)}_N\|_{\H^{(N)}}
\leq 2^p \|\ap\|_{\H^{(p)}}\,
\sum_{n=1}^\infty\left(\frac{4}{\hbar}\|w\|_\infty\,t\right)^n 
\end{align}
because $\ds \frac{(p+n)!}{p!n!}\leq 2^{n+p}$. The series on the R.S. of (\ref{est1}) converges for
$\ds |t|<\left(\frac{4}{\hbar}\|w\|_\infty\right)^{-1}$.  The third term in (\ref{Duhamel3})
is bounded similarly. Let
\be
A^{(p+n)}_{N,I}(i,j):=e^{iH^0_Nt_n/\hbar}e^{-iH_Nt_n/\hbar}A^{(p+n)}_N([g^{(n-1;p)}_{t_1,\ldots,t_{n-1}},w^{ij}_{t_n}])e^{iH^0_Nt_n/\hbar}e^{-iH_Nt_n/\hbar}
\ee
Then
\begin{align*}
& \frac1{N}\sum_{n=1}^\infty\bigg \|\sum_{i<j=1}^{p+n-1}\int_0^t\cdots\int_0^{t_{n-1}}
A^{(p+n)}_{N,I}(i,j)\,dt_n\cdots
dt_1\bigg\|_{\H^{(N)}} \leq
\\
&\frac1{N}\sum_{n=1}^\infty\frac{(p+n-1)^2}{2}\frac{2}{\hbar}\|w\|
\int_0^t\ldots\int_0^{t_{n-1}}\|A^{(p+n-1)}_N(g^{(n-1;p)}_{t_1,\ldots,t_{n-1}})\|_{\H^{(N)}}\,dt_n\ldots
dt_1\leq 
\\
& \leq 
\frac1{N}\sum_{n=1}^\infty(p+n-1)^2(p+n-2)\cdots
p\left(\frac{2}{\hbar}\|w\|_\infty\right)^{n-1}\frac{2}{\hbar}\|w\|_\infty
\|A^{(p)}\|_{\H^{(N)}}\frac{|t|^n}{n!}\leq
\\
& \leq
\frac{\|\ap\|_{\H^{(p)}}}{N}\sum_{n=1}^\infty(\frac{(p+n)!}{n!p!}\left(\frac{2}{\hbar}\|w\|_\infty
t\right)^{n}\leq 
\frac{\|\ap\|_{\H^{(p)}}}{N}2^p\sum_{n=1}^\infty
(\frac{4}{\hbar}\|w\|_\infty |t|)^{n} 
\end{align*}
Therefore
\begin{align*}
\bigg\|\frac1{N}\sum_{n=1}^k Q_{t,N}^{(p),n}\bigg\| _{\H^{(N)}}\leq 
\frac{\|\ap\|_{\H^{(p)}}}{N}2^p\sum_{n=1}^\infty (\frac{4}{\hbar}\|w\|_\infty |t!)^{n} .
\end{align*}
The remainder term in (\ref{Duhamel3}) clearly vanishes, as $k\to\infty$. To summarize,
 we have proven the following result.
\begin{proposition}. 
\label{conv1}
Let $\ds |t|<\left(\frac{4}{\hbar}\|w\|_\infty \right)^{-1}$. Then
\begin{align}
\label{exp1}
e^{iH_N t/\hbar} A^{(p)}_N e^{-iH_N t/\hbar}
=
A^{(p)}_{t}+\sum_{n=1}^\infty 
\int_0^t\ldots\int_0^{t_{n-1}}A^{(p+n)}_N(g^{(n;p)}_{t_1,\ldots,t_{n}})\,dt_n\ldots
dt_1 +O(1/N)
\end{align}
\end{proposition}
\subsection{Convergence for all times, $\hbar$ fixed}
We assume that the statement of Theorem \ref{mainth}
holds up to some time $T$
{\it independent} of $p$, i.e.,
\be
\label{T}
\limN \la \Psi_{N,0},e^{iH_N T/\hbar}A^{(p)}_N e^{-iH_N T/\hbar}\Psi_{N,0}\ra_{\H^{(N)}}=a^{(p)}(\psi_{p,T})
\ee
Let us proceed one step further in time, with $t<(4\|w\|_\infty/\hbar)^{-1}$. On account of (\ref{Duhamel3}), we have that
\begin{align}
\nonumber 
&
e^{iH_N (T+t)/\hbar}A_N^{(p)} e^{-iH_N (T+t)/\hbar}=
\\
\label{T1}
&
e^{iH_N T/\hbar}e^{- iH_Nt/\hbar}A_N^{(p)}
e^{- iH_Nt/\hbar}e^{-iH_N T/\hbar}=
e^{iH_N T/\hbar}A_{N}^{(p)}(t) e^{-iH_N T/\hbar}
\\
\nonumber
&+
\sum_{n=1}^N\,\int_0^t\ldots\int_0^{t_{n-1}}e^{iH_N T/\hbar}
A^{(p+n)}_N(g^{(n;p)}_{t_1,\ldots,t_{n}})e^{-iH_N
T/\hbar}\,dt_n\ldots dt_1+O(1/N)
\end{align}
This expansion is norm convergent, by (\ref{Duhamel3}) and the unitarity of 
$\ds e^{iH_N T/\hbar}$. Taking expectation values in $\Psi_{N,0}$ we get, as above,
\begin{align}
\nonumber
&
\la\Psi_{N,0},e^{iH_N (T+t)/\hbar}A_{N}^{(p)} e^{-iH_N (T+t)/\hbar}\Psi_{N,0}\ra_{\H^{(N)}}=
\\
\nonumber
&
=\la\Psi_{N,0},e^{iH_N T/\hbar}A_{t}^{(p)} e^{-iH_N T/\hbar}\Psi_{N,0}\ra_{\H^{(N)}}+
\\
\nonumber
&
\sum_{n=1}^{N}\,\int_0^t\ldots\int_0^{t_{n-1}}\la\Psi_{N,0},e^{iH_N
T/\hbar}
A^{(p+n)}_N(g^{(n;p)}_{t_1,\ldots,t_{n}})e^{-iH_N
T/\hbar}\Psi_{N,0}\ra_{\H^{(N)}}\,dt_n\ldots dt_1
\\
\nonumber
&+O(1/N)
\end{align}
Hence, by the inductive assumption and the norm convergence of the series
\begin{align*}
&
\limN\la\Psi_N,e^{iH_N (T+t)/\hbar}A_N^{(p)} e^{-iH_N (T+t)/\hbar}\Psi_N\ra_{\H^{(N)}}=
\ap(e^{i t\Delta/\hbar}\psi_T)+
\\
&
\sum_{n=1}^{\infty}\limN 
\,\int_0^t\ldots\int_0^{t_{n-1}}\la\Psi_{N,t},e^{iH_N
T/\hbar}
A^{(p+n)}_N(g^{(n;p)}_{t_1,\ldots,t_{n}})e^{-iH_N
T/\hbar}\Psi_{N,t}\ra_{\H^{(N)}}\,dt_n\ldots dt_1
\end{align*}
We postpone to Section 3, below, the proof that actually
\begin{align}
\nonumber
& \ap(e^{i t\Delta/\hbar}\psi_T)+
\\
\label{global}
&
\limN \sum_{n=1}^{\infty}
\,\int_0^t\ldots\int_0^{t_{n-1}}\la\Psi_{N,t},A^{(p+n)}_N(g^{(n;p)}_{t_1,\ldots,t_{n}})\Psi_{N,t}\ra_{\H^{(N)}}\,dt_n\ldots dt_1
\\
\nonumber
&
=\ap(\psi_{t+T }),
\end{align}
and this ensures that the convergence is global in time. 
\subsection{Control of the expansion, uniformity with respect to $\hbar$}
Given a symbol $\tau (x,\xi)\in \A_{\sigma,p}$, we denote by $T$ the corresponding Weyl operator.
Its action  on vectors  $\psi\in {\cal S}(\R^{3p})$ is given by
\be
\label{Weyl}
(T\psi)(x)=\frac1{\hbar^{3p}}\int_{\R^{3p}}\int_{\R^{3p}}\tau[(x+y)/2,\xi]e^{i\la
(x-y),\xi\ra/\hbar}\psi(y)\,dyd\xi
\ee
In general, $T$  is a semiclassical pseudodifferential operator. Let us recall some
relevant results (see e.g.\cite{Ro}).
\begin{enumerate}
\item
If $\widehat{\tau}\in L^1(\R^{3p}\times\R^{3p})$ then $T$
extends to a continuous operator on $L^2(\R^{3p})$ with $\ds \|T\|\leq \|\widehat{\tau}\|_{L^1}$;
hence 
$\ds \|T\|_{L^2\to L^2}\leq \|T\|_\sigma:=\|{\tau}\|_{\sigma}$, where $\widehat{\tau}$ is the Fourier transform of $\tau $, and
$$
\|{\tau}\|_{\sigma}:=\int_{\R^{6p}}|\hat{\tau}(s)|e^{\sigma |s|}\,ds .
$$
Obviously,
$\ds
\|\widehat{\tau}\|_{L^1}
\leq
\|\tau\|_{\sigma}$.
\item
If $w\in\As, g\in \As$, then the symbol of the commutator
$[W,G]/i\hbar$ is the Moyal bracket $\{w,g\}_M$.  Hence the multiple commutator
$[W,[W,\ldots [W,G]\ldots ]/(i\hbar)^n$ has as its symbol the multiple Moyal bracket 
$\{w,\{w,\ldots,\{w,g\}_M\ldots\}_M$. 
\newline
We recall that, given $g,g^{\prime}\in\As$, their Moyal
bracket $\{g,g^{\prime}\}_M$ is defined as
$$
\{g,g^{\prime}\}_M=g\# g^{\prime}-g^{\prime}\#g,
$$
where $\#$ is the composition of Weyl
symbols. In the Fourier transform representation the explicit expression of the Moyal bracket 
is (see e.g.
\cite{Fo},\S $3.4$): 
\be
\label{twisted}
(\{g,g^{\prime}\}_M)^{\wedge}(s)=
\frac{2}{\hbar}\int_{\R^{2n}}\widehat{g}(s^1)
\widehat{g^{\prime}}(s-s^1)
\sin{\left[{\hp}(s-s^1)\wedge s^1/{2}\right]}\,ds^1,
\ee
where, given two vectors $s=(v,w)$ and $s^1=(v^1,w^1)$, 
$s\wedge s^1:=\la w,v_1\ra-\la v,w_1\ra$.\par\noindent
\item If the observable $T$ has symbol $\tau(x,\xi)$, then the Heisenberg observable
$T_{t}$ has symbol $(\tau\circ \Phi_t^0)(x,\xi)$. Here $\Phi_t^0(x,\xi)=(x+\xi t,\xi)$
is the free flow with initial conditions $(x,\xi)$. 
 In particular, $(\tau\circ \Phi_t^0)(x,\xi)\in \As$ whenever $\tau\in \As$. 
\end{enumerate}
 Under the present assumptions, it can be proven, starting from the expression (\ref{twisted})
(see \cite{BGP}, Lemma 3.2 ), that   the following estimate on the Moyal bracket  holds:
\be
\label{M1}
\|\{w,g\}_M\|_{\sigma-\delta} \leq \frac{1}{e^2\delta^2}\|w\|_\sigma\,\|g\|_\sigma,
\quad 0<\delta<\sigma
\ee 
For the convenience of the reader  we reproduce here the proof of (\ref{M1}). 
Since  $(s-s^1)\wedge s^1=s\wedge s^1$, and $|s\wedge s^1|\leq |s|\cdot|s^1$, by definition of the  ${\mathcal A}_\sigma$-norm and (\ref{twisted}) we get:
\begin{align*}
&
\|(\{g,g^{\prime}\}_M)\|_{\sigma-\delta}=
\\
&
\frac{2}{\hbar}\int_{\R^{6p}}e^{(\sigma-\delta)|s|}\,ds \int_{\R^{6p}}|\hat{g}(s^1)\hat{g}^\prime(s-s^1)\sin{(\hbar(s-s^1)\wedge s^1)/2}|\,ds^1
\\
&
\leq \frac{2}{\hbar}\int_{\R^{6p}}\,ds \int_{\R^{6p}}e^{(\sigma-\delta)(|s|+|s^1|)}|\hat{g}(s)\hat{g}^\prime(s^1)\sin{(\hbar(s \wedge s^1)/2}|\,ds^1
\\
&
\leq \int_{\R^{6p}}e^{(\sigma-\delta)|s|}|\hat{g}(s)|\,ds \int_{\R^{6p}}e^{(\sigma-\delta)|s^1|}|\hat{g}^\prime(s^1)s \wedge s^1|\,ds^1
\\
&
\leq \int_{\R^{6p}}e^{(\sigma-\delta)|s|}|\hat{g}(s)||s|\,ds \int_{\R^{6p}}e^{(\sigma-\delta)|s^1|}|\hat{g}^\prime(s^1)s| |s^1||\,ds^1
\end{align*}
whence the assertion because  $\ds xe^{-\delta x}\leq \frac1{e\delta}$, $\forall\,x>0$, $\forall\delta>0$. 

Let now  $g_r:=\{g_{r-1},w\}_M, r>1; g_1=\{g,w\}_M$. Then, applying (\ref{M1}) $r$ times, we can write: 
\be
\label{M2}
\|g_r\|_{\sigma-r\delta}\leq
\left(\frac{1}{e^2\delta^2}\right)^r\|w\|_\sigma^r\,\|g\|_\sigma
\ee
These  results immediately yield the following bound.
\begin{lemma}. 
\label{lemma1}
Let the operator $\ap$ be the Weyl quantization of a symbol $\tau_a(x,\xi)\in {\cal A}_{\sigma,p}$ for some $\sigma >0$. Then there is $L(p)>0$ independent of $\hbar$ such that
\be
\label{unif1}
\|A^{(p+n)}_N(g^{(n;p)}_{t_1,\ldots,t_{n}})\|_{\H^{(N)}}\leq L^n\,n!^3\,(2\|w\|_\sigma)^n\|\ap\|_\sigma
\ee
\end{lemma}
{\bf Proof}

\noindent
Denote by ${\cal G}^{(n,p)}_{t,t_1,\ldots,t_{n}}$ the symbol of $g^{(n;p)}_{t_1,\ldots,t_{n}}$. Using
definition (\ref{treen}) and the estimate (\ref{M1}) we get the uniform estimate corresponding to
(\ref{normGn1}):
\be
\label{unif2}
\|{\cal G}^{(n,p)}_{t,t_1,\ldots,t_{n};N}\|_{\sigma-n\delta_n}\leq
\frac{2(p+n-1)}{e^2\delta_n^2}\|w\|_\sigma\,
\|{\cal G}^{(n-1,p)}_{t,t_1,\ldots,t_{n-1}}\|_\sigma , \;\;0<\delta_n<\sigma
\ee
The recursive definition (\ref{treen}) allows us to use the recursive estimate
(\ref{M2}).  We get
\begin{align}
\label{unif3}
\|{\cal G}^{(n,p)}_{t,t_1,\ldots,t_{n}}\|_{\sigma-n\delta_n} & \leq 2^n
({e^2\delta_n^2})^{-n}\frac{(p+n)!}{p!}\|w\|_\sigma^n\|a^{(p)}\|_{\sigma}
\end{align}
Setting $\ds \delta_n:=\frac{1}{2n}$  we get the bound (\ref{unif1}) on
account of the majorizations 
$$
 \|A^{(p+n)}_N(g^{(n;p)}_{t_1,\ldots,t_{n}})\|_{\H^{(N)}}\leq
\|{\cal G}^{(n,p)}_{t,t_1,\ldots,t_{n}}\|_{\sigma/2}, \quad \|w\|_{L^2\to L^2 }=\|w\|_\infty\leq
\|w\|_\sigma.
$$
 This
proves the Lemma.
\vskip 0.3cm\noindent
{\bf Remark} 
\newline
The uniform control in $\hbar$ introduces an extra $n!^2$ divergence
with respect to the fixed-$\hbar$ estimate (\ref{normGn1}).
\vskip 0.3cm\noindent
We now obtain  uniform estimates of the three terms in expansion
(\ref{Duhamel3}). 
\begin{lemma}.
\label{lemma2}
There exist constants $M_1>0, M_2>0, M_3>0, L_1>0, L_2>0, L_3>0$, independent of $(\hbar,t)$ and
$N$, such that
\begin{eqnarray}
\label{unif4}
\|B_{t,N}^{(p),k}\|_{\H^{(N)}}&\leq& M_1\|\ap\|_\sigma\sum_{n=1}^kL_1^n\,n!^2(\|w\|_\sigma t)^n
\\
\label{unif5}
\|Q_{t,N}^{(p),n} \|_{\H^{(N)}}&\leq& M_3\|\ap\|_\sigma L_2^n\,n!^2(\|w\|_\sigma\,t)^n
\\
\label{unif6}
\|R_{t,N}^{(p),k}\|_{\H^{(N)}} &\leq& M_3\|\ap\|_\sigma L_3^k\,k!^2(\|w\|_\sigma\,t)^k
\end{eqnarray}
\end{lemma}
{\bf Proof}

\noindent
Inserting the estimate (\ref{unif1}) in the expressions
(\ref{Duhamel4},\ref{Duhamel5}, \ref{Duhamel6}) we get, on account of unitarity of 
$U_0(t)$: 
\begin{eqnarray*}
\|B_{t,N}^{(p),k}\|_{\H^{(N)}}&\leq &
\|\ap\|_\sigma\sum_{n=1}^{k}\,(2L\|w\|_\sigma)^n\,n!^3\,\int_0^t\ldots\int_0^{t_{n-1}}\,dt_n\ldots dt_1
\\
&\leq & \|\ap\|_\sigma \,\sum_{n=1}^{k}\,(2L\|w\|_\sigma |t|)^n\,n!^2.
\end{eqnarray*}
 The last inequality comes from performing the time integrations, which are majorized 
by a factor
$|t|^n/n!$ in (\ref{unif4},\ref{unif5}) and by a factor
$|t|^k/k!$ in (\ref{unif6}) (proven with the help of the same argument). This proves the lemma.
\vskip 0.3cm\noindent
Using this result, we can easily prove the uniform version of the
expansion (\ref{Duhamel3}).
\begin{proposition}.
\label{repre} Let $\epsilon:=\|w\|_\infty t$. Then, in the same assumption of Lemma \ref{lemma1} on the operator $\ap$,  there exists $k=k(\ep)$, $\Lambda=\Lambda(\ep)$
such that
\begin{eqnarray}
\label{rep1}
e^{iH_Nt/\hbar}A^{(p)}_Ne^{-iH_Nt\hbar}&=&A^{(p)}_{t,N}+B_{t,N}^{(p),k}+ R_{t,N}^{(p),k} +\frac{\Lambda}{N},
\end{eqnarray} 
where
\be
\label{unifbis}
B_{t,N}^{(p),k}=\sum_{n=1}^k\int_0^t\ldots\int_0^{t_{n-1}}A^{(p+n)}_N(g^{(n;p)}_{t_1,\ldots,t_{n}})\,dt_n\ldots 
dt_1.
\ee
Here $B_{t,N}^{(p),k}$ fulfills the majorization (\ref{unif4}), and 
\be
\label{unif7}
\|R_{t,N}^{(p),k}\|_{\H^{(N)}}\leq M_3e^{-L_3/\sqrt{\epsilon}}
\ee
\end{proposition}
{\bf Proof}

\noindent
The estimate (\ref{unif5}) and a standard Nekhoroshev-type argument show that the choice 
\be
\label{kopt}
 k(\ep):=\frac1{\sqrt{\ep}}=\frac{1}{\|w\|_\infty\,t}
\ee
 minimizes the divergence of $R_{t,N}^{(p),k}$. A straightforward computation then yields (\ref{unif7}). By definition of $Q^{(p),N}_{t,N}$ we get the uniform version of the estimate (2.21), whence
$$
\Lambda(\ep):=p2^p\|\ap\|_{\sigma}\sum_{n=1}^{k(\ep)}n!^2(2\ep)^n\leq p2^p\ep^{-1/2}(e/2)^{-1/\sqrt{\ep}}
$$
\vskip 1.0cm\noindent
\begsection{Connection with the Hartree equation and proof of the theorems}
We wish to prove that the representation of the evolution obtained in Proposition \ref{conv1} 
  coincides   with the evolution generated by
the Hartree equation in the limit $N\to\infty$ .
  
\noindent
For this purpose, we recall that the Hartree equation is Hamiltonian.  We define the functional
\be
\label{Ham}
{\cal H}(\psi,\overline\psi)=-\frac{\hbar^2}{2} \int_{\R^3}|\nabla\psi(x)|^2\,dx+
{\cal W}(\psi,\overline\psi),
\ee
for $\psi\in H^{1}(\R^{3})$, where
\be
{\cal W}(\psi,\overline\psi)=\frac12\int_{\R^3\times\R^3}\psibx\psiby   
w(x-y)\psi(x)\psi(y)\,dxdy
\ee
If $\psi(x), \psiby$ are considered as canonical variables with Poisson brackets
$$
\{\psi(x),\psiby)\}=i\hbar\delta(x-y), \quad \{\psi(x),\psi
(y)\}=
\{\psibx,\psiby\}=0,
$$
then (\ref{Ham}) is the Hamiltonian functional generating a time evolution of 
fun\-ctionals on phase space equivalent to the Hartree equation. Namely, if
${\cal A}(\psi)$ is a functional and ${\cal A}_t$ denotes its time evolution, one has that
$$
\partial_t{\cal A}_t(\psi)=\frac{1}{\hbar}\{{\cal H},{\cal A}_t\}(\psi)
$$
Choosing ${\cal A}=\langle\phi,\psi\rangle$, $\phi\in C_0^\infty(\R^3)$,
then    ${\cal A}_t(\psi)={\cal A}(\psi_t)$, where $\psi_t$ is a solution of the Hartree equation
\be
\label{Har2}
i\hbar\partial_t \psi_{t}=-\frac{\hbar^2}{2}\Delta\psi_{t}+(w\ast|\psi_{t}|^{2})\psi_{t}
\ee
\noindent
Define the free flow $\Phi_t^0({\cal A}):={\cal A}_t$ of ${\cal A}$ by
$$
{\cal A}_t={\cal A}(e^{i\Delta t/\hbar}\psi)
$$
and denote by $\Phi_t({\cal A})$ the interacting flow.  Formally, the interacting flow is given by
the Lie expansion in the interaction representation (analogous to the Schwinger-Dyson expansion of Section 2.2). Indeed we have the following result: 
\par\noindent
\begin{lemma}.
\label{lLie}
$\Phi_t({\cal A})$ {\it admits the formal expansion}
\be
\label{Lie}
\Phi_t({\cal A})={\cal A}_t+\Phi^0_t\left(\sum_{n=1}^\infty \left(\frac1{\hbar}\right)^n\int_0^t\ldots
 \int_0^{t_n}\{ {\cal W}_{t_n}\ldots
\{{\cal W}_{t_1},{\cal A}\}\ldots\}\,dt_n\ldots dt_1\right)
\ee
\end{lemma}
{\bf Proof.} To see this, we consider the dynamics in the interaction picture. We set
$$
\tilde{{\cal A}}_t := \Phi_t\circ \Phi^0_{-t}({\cal A})
$$
Then
$$
\partial_t \tilde{{\cal A}}_t=\frac{\{\tilde{\cal A}_t,{\cal W}_{t}\}}{\hbar}
$$
where ${\cal W}_t:=\Phi^0_t({\cal W})$ is the free evolution of ${\cal W}$. After integrating in time we get
$$
\tilde{{\cal A}}_t={\cal A}+ \int_0^t\frac{\{{\cal W}_{s},{\cal A}\}}{\hbar}\,ds
$$
whence
$$
\Phi_t({{\cal A}})={\cal A}_t+ \Phi^0_t\left(\int_0^t\frac{\{{\cal W}_{s},{\cal A}\}}{\hbar}\,ds\right)
$$
Iterating this identity, we obtain the series (\ref{Lie}), and this concludes the proof of the Lemma. 
\par\noindent
The desired identification is based on the following proposition
\begin{proposition}.  
Let $\psi\in H^1(\R^3)$, and let $\Psi$ be a product state, i.e.
$$
\Psi(x_1,\ldots,x_l)=\prod_{s=1}^l\psi(x_s)
$$
Then, for all $N\geq p$,
\begin{align}
\label{equality}
g^{(n;p)}_{t_1,\ldots,t_{n}}(\psi)&:=\frac{(p+n)^{p+n}}{(p+n)!}\la\Psi_{n+p},A^{(p+n)}_{p+n}(g^{(n;p)}_{t_1,\ldots,t_{n}})\Psi_{n+p}\ra_{\H^{(n+p)}} =
\\
\nonumber
& \left(\frac1{\hbar}\right)^n\{{\cal W}_{t_n}\ldots\{ {\cal W}_{t_1},{\cal A}\}\ldots\}(\psi), 
\end{align}
where
$$
{\mathcal A}=\ap(\psi):=\int\,\overline{\psi(x_1)}\cdots\overline{\psi(x_p)}\alpha^{(p)}(x_1,\ldots,x_p;y_1,\ldots,y_p)\overline{\psi(y_1)}\cdots\overline{\psi(y_p)}\prod_{k=1}^p\,dx_kdy_k
$$
\end{proposition}
{\bf Proof.}
\noindent
We have that
$$
\left.\Phi_t\circ
\Phi^0_{-t}(\{{\cal W}_t,{\cal A}\}/\hbar)\right|_{t=0}=\{{\cal W},{\cal A}\}/\hbar=\partial_t\tilde{{\cal A}}|_{t=0}
$$
Define the projection $\rho:=|\psi\rangle\langle\psi|$.  Then, denoting $\tilde{\rho}_t:=\Phi_t\circ \Phi^0_{-t}({\rho})$,  we have  that, 
for all $n\geq 1$, 
$$
\tilde{{\cal A}}_t={\rm Tr}({\cal A}\tilde{\rho}_t^{\otimes n});\quad \partial_t\tilde{{\cal A}}|_{t=0}=\partial_t{\rm Tr}({\cal A}\tilde{\rho}_t^{\otimes n})|_{t=0}.
$$
Therefore, in the interaction picture
$$
\partial_t\tilde{\rho}_t={\rm Tr}(\tilde{\rho}_t^{\otimes 2}{\cal W}^{12}_t-
{\cal W}^{12}_t\tilde{\rho}_t^{\otimes 2})/\hbar
$$
and in the same way we get
$$
\frac1{\hbar}\{ {\cal W}_t, {\cal A}\}=\sum_{i=1}^n{\rm Tr}(({\cal W}_t^{in+1}{\cal A}-{\cal A}{\cal W}_t^{in+1})\rho^{\otimes\, n+1})
$$
It is then easy to check that
\begin{eqnarray*}
\frac1{\hbar^n}\{{\cal W}_{t_n}\ldots\{ {\cal W}_{t_1},{\cal A}\}\ldots\}(\psi)&=&{\rm Tr}(g^{(n;p)}_{t_1,\ldots,t_{n}}
\cdot \rho^{\otimes\, n+p})
\\
&=&\frac{(p+n)^{p+n}}{(p+n)!} \langle\Psi_{n+p},A^{(p+n)}_{p+n}(g^{(n;p)}_{t_1,\ldots,t_{n}})\Psi_{n+p}\rangle_{\H^{(n+p)}}
\\
&=&
g^{(n;p)}_{t_1,\ldots,t_{n}}(\psi),
\end{eqnarray*}
and this concludes the proof of the Proposition.
\vskip 0.3cm\noindent
We are now in a position to prove our main results.
\par\noindent
{\bf Proof of Theorem \ref{mainth}}

\noindent
Consider the expectation value of the expansion (\ref{exp1}) in a coherent (i.e., product) state:
\begin{align*}
& \la \Psi_N,e^{iH_Nt/\hbar}A_N^{(p)}e^{-iH_Nt/\hbar}\Psi_N\ra_{\H^{(N)}}=\la\Psi_N,A_{t,N}^{(p)}\Psi_N\ra_{\H^{(N)}}+
\\
&\sum_{n=1}^{N}\int_0^t\ldots\int_0^{t_{n-1}}\la\Psi_N,A^{(p+n)}_N(g^{(n;p)}_{t_1,\ldots,t_{n}})\Psi_N\ra_{\H^{(N)}}\,dt_n
\cdots dt_1+O(1/N).
\end{align*}
By definition of $\Psi_N$, 
\begin{align*}
&\frac{N^p}{N(N-1)\ldots (N-p+1)}\la \Psi_N,e^{iH_Nt/\hbar}A_Ne^{-iH_Nt/\hbar}\Psi_N\ra_{\H^{(N)}}
=\la\Psi_p,a_t^{(p)}\Psi_p\ra_{\H^{(p)}}+
\\
& +\sum_{n=1}^{\infty}\int_0^t\ldots\int_0^{t_{n-1}}
\la\Psi_{n+p},g^{(n;p)}_{t_1,\ldots,t_{n}}\Psi_{n+p}\ra_{\H^{(n+p)}}\,dt_n\cdots dt_1+O(1/N).
\end{align*}
Since the series is norm- convergent, the limits 
$N\to\infty$ and $n\to\infty$ can be interchanged. Then
\begin{eqnarray}
\label{final}
\limN
\la\Psi_N,e^{iH_Nt/\hbar}A_Ne^{-iH_Nt/\hbar}\Psi_N\ra_{\H^{(N)}}=
\la\Psi_p,a_t^{(p)}\Psi_p\ra_{\H^{(p)}}+\qquad\qquad\\
\nonumber
\sum_{n=1}^{\infty}\int_0^t\ldots\int_0^{t_{n-1}}
\la\Psi_{n+p},g^{(n;p)}_{t_1,\ldots,t_{n}}\Psi_{n+p}\ra_{\H^{(n+p)}}\,dt_n\cdots dt_1=
\ap(\psi_t),
\end{eqnarray}
where the last equality follows from formula (\ref{equality}). 
\vskip 0.3cm\noindent
{\bf  Proof of Theorem \ref{unif}}

\noindent
If, instead of (\ref{exp1}), the  representation (\ref{rep1}) is considered, the above argument yields
\begin{align}
\label{final1}
&\frac{N^p}{N(N-1)\ldots (N-p+1)}\la
\Psi_N,e^{iH_Nt/\hbar}A_N^{(p)}e^{-iH_Nt/\hbar}\Psi_N\ra=
\la\Psi_p,a^{(p)}_t\Psi_p\ra+\la\Psi_N,R^{k,p}_{t,N}\psi_N\ra
\\
\nonumber
&+\sum_{n=1}^{k(\ep)}\int_0^t\ldots\int_0^{t_{n-1}}
\la\Psi_{n+p},g^{(n;p)}_{t_1,\ldots,t_{n}}\Psi_{n+p}\ra\,dt_n\cdots dt_1=
a^{(p)}(\psi_t)+O(e^{-1/\sqrt{\epsilon}}).
\end{align}
Given any  bounded operator $A$ on $L^2(\R^{3l})$ with (Weyl) symbol  $\sigma_A(x,\xi): {\cal S}(\R^{6l})\to\R$,   where ${\mathcal S}$ is the Schwartz space of rapidly decreasing functions, its matrix elements can be expressed in terms of the symbol and of the Wigner function by the following well known formula (see e.g.\cite{Fo}):
\be
\label{W4}
\la \Psi,A\Psi\ra=\int_{\R^{3l}\times\R^{3l}}\sigma_A(x,\xi)W_\Psi(x,\xi)\, dxd\xi
\ee
where $W_\Psi(x,\xi)$ is the Wigner function of the state $\Psi$. Therefore, in our case
$$
\frac{N^p}{N(N-1)\ldots (N-p+1)}\la \Psi_N,A_N(t)\Psi_N\ra=\int_{\R^{3p}\times\R^{3p}}\sigma_A(X_p,\Xi_p)W_N^{\Psi_N}(X_p,\Xi_p,t)\, dX_pd\Xi_p, 
$$
where $\ds W_N^{\Psi_N}(X_p,\Xi_p,t)$ is the Wigner function corresponding to the time evolution, \linebreak
$\ds e^{iH_N t/\hbar}\Psi_N$, of the product state $\Psi_{N,0}=\psi(x_1)\ldots\psi(x_N)$. The $N-p$ variables $(X_{N-p},\Xi_{N-p})$ are integrated out. By 
 (\ref{final1})  
and (\ref{W3}), we
can take the $N\to\infty$ limit and write
\begin{eqnarray*}
\int_{\R^{3p}\times\R^{3p}}\sigma_A(X_p,\Xi_p)W_N^{\Psi_N}(X_p,\Xi_p,t)\, dX_pd\Xi_p=
\\
\int_{\R^ {3p} \times   \R^ {3p} } \, 
\sigma_A(X_p,\Xi_p)
\prod_{l=1}^p 
W_{\psi}(x_l,\xi_l;t) 
\, dX_pd\Xi_p
+O(e^{
-1/\sqrt{\ep}}).
\end{eqnarray*}
Since this formula holds for any $\sigma_A(X_p,\Xi_p)\in {\cal
S}(\R^{3p}\times\R^{3p})\cap {\mathcal A}_{\sigma,p}$, the assertion is proved.
\par\noindent
{\bf Proof of formula (\ref{global})}.
\par\noindent
By (\ref{equality}),  we have that
\begin{align*}
&
\la\Psi_N,e^{iH_N
T/\hbar}
A^{(p+n)}_N(g^{(n;p)}_{t_1,\ldots,t_{n}})e^{-iH_N
T/\hbar}\Psi_N\ra_{\H^{(N)}}
=
\\
&
\left(\frac1{\hbar}\right)^n\{{\cal W}_{t_n}\ldots\{{\cal W}_{t_1},{\cal A}\}\ldots\}(\psi_T)
\end{align*}
which yields formula (\ref{global}), by Lemma \ref{lLie},  on account of the uniform convergence of the series. 
\vskip 1.0cm\noindent

\vfill\eject

\end{document}